\documentclass{revtex4}
\usepackage{epsf}

\begin{document}
\title{Negative parity states of $^{11}$B and $^{11}$C
 and the similarity with 
$^{12}$C}

\author{Yoshiko Kanada-En'yo}
\address{Yukawa Institute for Theoretical Physics, Kyoto University,\\
Kyoto 606-8502, Japan}

\begin{abstract}
The negative parity states of $^{11}$B and $^{11}$C were studied
based on the calculations of antisymmetrized
molecular dynamics(AMD). The calculations well reproduced 
the experimental strengths of Gamov-Teller(GT), 
$M1$ and monopole transitions. 
We, especially, focused on the $3/2^-_3$ and $5/2^-_2$ states, for which
GT transition strengths were recently measured.
The weak $M1$ and GT transitions for the $3/2^-_3$ in $^{11}$B and 
$^{11}$C are described by
a well-developed cluster 
structure of $2\alpha$+$t$ and $2\alpha$+$^3$He, respectively,
while the strong transitions for the $5/2^-_2$ is characterized by 
an intrinsic spin excitation with no cluster structure.
It was found that the $3/2^-_3$ state is a dilute cluster state, and
its features are similar to those of 
the $^{12}$C$(0^+_2)$ which is considered to
be a gas state of three $\alpha$ clusters.
\end{abstract}
\maketitle

\noindent
%PACS numbers: 21.60.-n, 02.70.Ns, 21.10.Ky, 27.20.+n.

\section{Introduction}	

Cluster aspect is known to be one of essential features
in light nuclei.
Recently, various new types of cluster structure have been predicted and
found in excited states of light stable nuclei 
as well as in light unstable nuclei. 
In case of $^{12}$C, it was known that 
3$\alpha$-cluster states develop in such excited states as 
the $0^+_2$(7.65 MeV) state. 
Tohsaki {\it et al.}\cite{Tohsaki01,Funaki03} 
proposed a new interpretation of the $0^+_2$ as 
a dilute gas state of weakly interacting 3 $\alpha$ particles.
It is challenging problem to answer the question whether or not
such a cluster gas is the general feature which appears in
other nuclear systems.
In order to search for such the dilute cluster states,   
we studied the structure of excited states of $^{11}$C and $^{11}$B.

The present study has been motivated by the recent measurements of 
Gamov-Teller (GT) transitions $^{11}$B$\rightarrow ^{11}$C$^*$
with high energy resolutions\cite{Kawabata04,Fujita04}.
In the experiments, the GT transition strengths to the $3/2^-_3$ and
the $5/2^-_2$ states were separately measured, and 
the transition to the $^{11}{\rm C}(3/2^-_3,8.10 {\rm MeV})$
was found to be extremely weak compared with that to the
$^{11}{\rm C}(5/2^-_2,8.42 {\rm MeV})$ and also with those to 
other low-lying states.
Abnormal features of the $3/2^-_3$ have been known also in the mirror
nucleus $^{11}$B. For example, the $3/2^-_3$ of $^{11}$B has relatively 
weak $M1$ transitions into the lower states compared with strong 
transitions among other low-lying states. Another characteristic of the
$^{11}$B($3/2^-_3$) is the strong monopole transition
observed by the recent experiments of the inelastic $(d,d')$ scattering,
where similarities of the $^{11}$B($3/2^-_3$) with the
$^{12}$C($0^+_2$) were suggested\cite{Kawabata05}.
In the theoretical side,
the structure of the $^{11}$B($3/2^-_3$) state has been mysterious 
because this state can not be described by any models.
No theoretical state can be assigned to the
$3/2^-_3$ in shell model calculations\cite{Cohen65,Gomez93,navratil03} nor 
cluster model calculations\cite{Nishioka79}.
These facts indicate that the $3/2^-_3$ of $^{11}$B and $^{11}$C 
may have an abnormal structure,
and is a candidate of the dilute cluster state. 
On the other hand, the shell models succeeded to reproduce  
various properties of the low-lying negative-parity states
with the excitation energy $E_x < 9$ MeV except for 
the $3/2^-_3$\cite{navratil03}.
It suggests the possible coexistence of cluster states and non-cluster states 
in $^{11}$B and $^{11}$C.

In this paper, we study the negative-parity states 
of $^{11}$B and $^{11}$C based on the theoretical calculations of
antisymmetrized molecular dynamics(AMD). We apply the method of variation
after spin-parity projections in the AMD framework, which has been proved to
be a powerful tool for studying excited states of light nuclei.
We focus on the structure of the $3/2^-_3$ and the $5/2^-_2$ 
around $E_x=8$ MeV, and show the similarity of 
the excited states of $^{11}$B with those of $^{12}$C.

The paper is organized as follows. First, we briefly explain the theoretical
method in \ref{sec:formulation}, 
and then we show the calculated results comparing
with the experimental data in \ref{sec:results}.
In \ref{sec:discuss}, the structure of excited states of $^{11}$B 
is discussed, and their similarity with $^{12}$C is shown.
Finally, we give a summary in \ref{sec:summary}.

\section{Formulation} \label{sec:formulation}
We perform the energy variation after 
spin parity projection(VAP) within the AMD model space, as was done
in the previous studies\cite{Enyo-c12,Enyo-be10}. 
The detailed formulation of the AMD method 
for nuclear structure studies is described in 
\cite{Enyo-c12,Enyo-be10,ENYObc,ENYOsup,AMDrev}.
In particular, the formulation of the present calculations is basically 
the same as that described in \cite{Enyo-c12,Enyo-be10,Enyo-c12v2}.

An AMD wave function is a Slater determinant of Gaussian wave packets;
\begin{equation}
 \Phi_{\rm AMD}({\bf Z}) = \frac{1}{\sqrt{A!}} {\cal{A}} \{
  \varphi_1,\varphi_2,...,\varphi_A \},
\end{equation}
where the $i$-th single-particle wave function is written by a product of
spatial($\phi$), intrinsic spin($\chi$) and isospin($\tau$) 
wave functions as,
\begin{eqnarray}
 \varphi_i&=& \phi_{{\bf X}_i}\chi_i\tau_i,\\
 \phi_{{\bf X}_i}({\bf r}_j) &\propto& 
\exp\bigl\{-\nu({\bf r}_j-\frac{{\bf X}_i}{\sqrt{\nu}})^2\bigr\},
\label{eq:spatial}\\
 \chi_i &=& (\frac{1}{2}+\xi_i)\chi_{\uparrow}
 + (\frac{1}{2}-\xi_i)\chi_{\downarrow}.
\end{eqnarray}
$\phi$ and $\chi$ are represented by 
complex variational parameters, ${\rm X}_{1i}$, ${\rm X}_{2i}$, 
${\rm X}_{3i}$, and $\xi_{i}$. The iso-spin
function $\tau_i$ is fixed to be up(proton) or down(neutron). 
We use the fixed width parameter $\nu=0.19$ fm$^{-2}$, which 
is chosen to be the optimum value for $^{11}$B.
Accordingly, an AMD wave function
is expressed by a set of variational parameters, ${\bf Z}\equiv 
\{{\bf X}_1,{\bf X}_2,\cdots, {\bf X}_A,\xi_1,\xi_2,\cdots,\xi_A \}$.

For the lowest $J^\pi$ state,
we vary the parameters ${\bf X}_i$ and $\xi_{i}$($i=1\sim A$) to
minimize the energy expectation value of the Hamiltonian,
$\langle \Phi|H|\Phi\rangle/\langle \Phi|\Phi\rangle$,
for the spin-parity projected AMD wave function;
$\Phi=P^{J\pi}_{MK'}\Phi_{\rm AMD}({\bf Z})$.
Here, $P^{J\pi}_{MK'}$ is the spin-parity projection operator.
Then we obtain the optimum solution of the parameter set;
${\bf Z}^{J\pi}_1$ for the lowest $J^\pi$ state.
The solution ${\bf Z}^{J\pi}_n$ 
for the $n$-th $J^\pi$ state are calculated by varying ${\bf Z}$ 
so as to minimize the energy of the wave function; 
\begin{equation}
|\Phi\rangle =|P^{J\pi}_{MK'}\Phi_{\rm AMD}({\bf Z})\rangle
-\sum^{n-1}_{k=1}
|P^{J\pi}_{MK'}\Phi_{\rm AMD}({\bf Z}^{J\pi}_k)\rangle 
\frac{
\langle P^{J\pi}_{MK'}\Phi_{\rm AMD}({\bf Z}^{J\pi}_k)
||P^{J\pi}_{MK'}\Phi_{\rm AMD}({\bf Z})\rangle}
{\langle P^{J\pi}_{MK'}\Phi_{\rm AMD}({\bf Z}^{J\pi}_k)|P^{J\pi}_{MK'}\Phi_{\rm AMD}({\bf Z}^{J\pi}_k)\rangle},
\end{equation}
which is the orthogonal component to the lower states. 

After the VAP calculations of the $J^\pi_n$ states for various
$J$, $n$ and $\pi=\pm$,
we obtained the optimum intrinsic states,
$\Phi_{\rm AMD}({\bf Z}^{J\pi}_n)$, 
which approximately describe the corresponding $J^\pi_n$ states. 
In order to improve the wave functions, we superpose all the 
obtained AMD wave functions. 
Namely, we determine
the final wave functions for the $J^\pi_n$ states as, 
\begin{equation}\label{eq:diago}
|J^\pi_n\rangle=\sum_{i,K} c(J^\pi_n,i,K) 
|P^{J\pi}_{MK}\Phi_{\rm AMD}({\bf Z}^{J_i\pi_i}_{k_i})\rangle,
\end{equation}
where the coefficients $c(J^\pi_n,i,K)$ are determined by the 
diagonalization of the Hamiltonian and norm matrices.
Here the number of the independent AMD wave functions, which are superposed
in Eq.\ref{eq:diago}, is that of the spin parity states $\{J^\pi_n\}$ 
calculated by the VAP.
We calculate the expectation values for various observables
with the $|J^\pi_n\rangle$ obtained after the diagonalization.

\section{Results} \label{sec:results}

We adopt the same effective nuclear interaction as those used
in Ref.~\cite{Enyo-c12}, which consists of the central force, the
spin-orbit force and the Coulomb force.
The interaction parameters are slightly modified from the previous ones 
for better reproduction of the energy levels of $^{11}$B and $^{11}$C. 
Namely, the Bartlett, Heisenberg and Majorana parameters in the MV1 force
are chosen to be $b=h=0.25$ and $m=0.62$, and the 
strengths of the spin-orbit force are $u_{I}=-u_{II}=2800$ MeV.

The base AMD wave functions are obtained by the VAP for the ground and
the excited states of $^{11}$B. The number of the base AMD wave functions
in the present calculations are 17. These independent AMD wave functions are
superposed to calculate the final wave functions.
In the calculations of the $^{11}$C,
we assume the mirror symmetry of the base AMD wave functions for simplicity.
The coefficients of the base wave functions in the superposition are optimized
for each system of $^{11}$B and $^{11}$C. 
 
\begin{figure}[th]
\epsfxsize=8. cm
\centerline{\epsffile{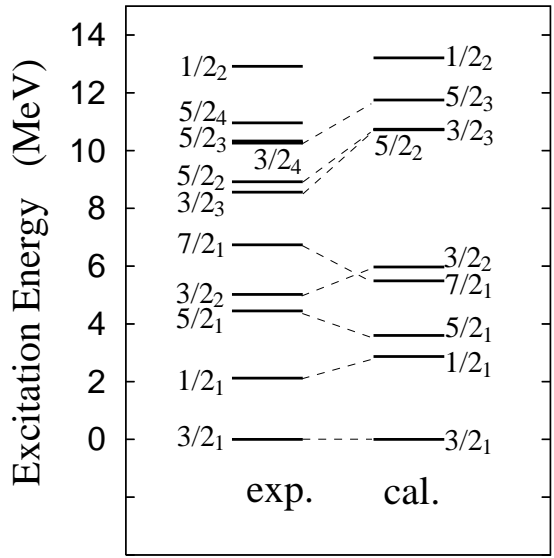}}
\vspace*{8pt}
\caption{Energy levels of the negative-parity states of $^{11}$B.
\protect\label{fig:b11spe}
}
\end{figure}

 The energy levels of the negative parity states in $^{11}$B are shown 
in Fig. \ref{fig:b11spe}. In the results, we obtain the $3/2^-_3$ and 
the $5/2^-_2$ at about $E_x=10$ MeV. 
We can assign the obtained 
$3/2^-_3$ and $5/2^-_2$ to the observed $3/2^-_3$($E_x=8.56$ MeV) and
$5/2^-_2$($E_x=8.92$ MeV) due to the good agreements 
of transition strengths between theory and
experimental data as shown later,
though the excitation energies are overestimated 
by the present calculations.

%The properties of the ground states of $^{11}$B and $^{11}$C  
%such as $\mu$-moments and $Q$-moments are well reproduced by theory
%as shown in table \ref{tab:bgt}. 

The GT transition strengths from the $^{11}$B$_{\rm g.s.}$ to $^{11}$C$^*$,
the $M1$ and $E2$ transition strengths in $^{11}$B are shown in tables
\ref{tab:bgt}, \ref{tab:bm1} and \ref{tab:be2},respectively. 
The calculated 
values for these transitions are in good agreements with the observed values.
The $B({\rm GT})$ for the transitions to the $^{11}$C$^*(3/2^-_3)$ and 
the $^{11}$C$^*(5/2^-2)$ at $E_x\sim 8$ MeV were recently measured by charge 
exchange reactions\cite{Kawabata04,Fujita04}, and it was found that 
$B({\rm GT};^{11}$B$\rightarrow ^{11}$C($3/2^-_3$)) is abnormally small
while $B({\rm GT};^{11}$B$\rightarrow ^{11}$C($5/2^-_2$)) is as large as
those for other low-lying states of $^{11}$C.
The present result well describes 
the small $B({\rm GT})$ for the $3/2^-_3$ state because
it has a well developed $2\alpha+^3$He cluster structure, 
and hence, the structure of the daughter state much differs from the 
normal structure of the parent state, $^{11}$B$_{\rm g.s.}$. 
Due to the same reason, the $M1$ transitions from the 
$^{11}$B$(3/2^-_3)$ to the low-lying states are generally weak compared with
other $M1$ transitions among the low-lying states.
On the other hand, since the $5/2^-_2$ has no cluster structure, 
the $B({\rm GT})$ for the $^{11}$C$^*(5/2^-_2)$ 
and the $B(M1)$ for the $^{11}$B$^*(5/2^-_2)$ are as large as those
for the other low-lying states in the theoretical results. This is consistent
with the experimental data. We also show the theoretical 
$B({\rm GT})$ calculated by the no-core shell model(NCSM)\cite{navratil03} 
in table \ref{tab:bgt}. The strengths of the GT transitions to $^{11}$C$^*$
are reproduced also by NCSM except for the transition 
to the $^{11}$C$^*(3/2^-_3)$. In the NCSM, 
 the $3/2^-_3$ can not be described 
because the limited model space in the shell model is not suitable 
to describe cluster states with the spatial development.

In the recent experiments of the inelastic $(d,d')$ scattering
\cite{Kawabata05}, 
it has been found that the iso-scalar monopole transition 
for $3/2^-_1\rightarrow 3/2^-_3$ is as strong as $B(E0;IS)=94\pm 16$
fm$^4$. The calculated strength for this inelastic transition is 
$B(E0;IS)=94$ fm$^4$, which well agrees with the experimental data.

\begin{table}[ht]
\caption{ 
\protect\label{tab:bgt} Comparison of the GT transition strengths,
the binding energies, $Q$-moments, $\mu$-moments and $2\alpha-t$ threshold
between present results and the experimental data.
The theoretical values obtained by the no-core shell model 
calculation with AV8'+TM'(99) \protect\cite{navratil03} 
are also shown.
%The adopted interaction consists of
%the central force of MV1 ($m=0.62$, $b=h=0.25$),
%the spin-orbit force of GSRS ($u_I=-u_{II}=2800$ MeV), and Coulomb force. 
The experimental data are taken from Ref.\protect\cite{Fujita04}.
}{
\begin{tabular}{c@{\qquad\qquad}c@{\qquad}c@{\qquad}c@{\qquad}} 
%\toprule
%\begin{tabular}{@{}ccccc@{}} \toprule
%{ccccc}
%hline\hline
 & exp. & AMD & NCSM  \\ 
%\colrule
\hline
$B({\rm GT};^{11}$B$\rightarrow ^{11}$C($3/2^-_1$))  &  0.345(8) &   0.40 & 0.315   \\
$B({\rm GT};^{11}$B$\rightarrow ^{11}$C($1/2^-_1$))  &  0.440(22) &  0.43 & 0.591    \\
$B({\rm GT};^{11}$B$\rightarrow ^{11}$C($5/2^-_1$))  &  0.526(27) &  0.70 &0.517    \\
$B({\rm GT};^{11}$B$\rightarrow ^{11}$C($3/2^-_2$))  &  0.525(27) &  0.67 & 0.741    \\
$B({\rm GT};^{11}$B$\rightarrow ^{11}$C($3/2^-_3$))  &  0.005(2) &   0.02 &    \\
$B({\rm GT};^{11}$B$\rightarrow ^{11}$C($5/2^-_2$))  &  0.461(23) &  0.56  & 0.625  \\
\hline
B.E.$(^{11}$B$_{\rm g.s.})$ [MeV]  &   76.205 &  72.8  & 73.338 \\
$\mu(^{11}$B$_{\rm g.s.})$ [$\mu^2_N$] & +2.689   & +2.3 & +2.176 \\
$Q(^{11}$B$_{\rm g.s.})$ [$e$ fm$^2$] & +4.065(26)   & +4.7 & +2.920 \\
B.E.$(^{11}$C$_{\rm g.s.})$ [MeV]     &  73.440  & 70.4 & 70.618 \\
$\mu(^{11}$C$_{\rm g.s.})$ [$\mu^2_N$]  &  $-0.964$  & $-0.6$ & $-0.460$ \\
$Q(^{11}$C$_{\rm g.s.})$ [$e$ fm$^2$] & +3.327(24) & +3.8 & +2.363\\
$2\alpha+t$ threshold [MeV] & 65.07 & 70.6 & \\
\end{tabular}}
\end{table}

\begin{table}[ht]
\caption{ 
\protect\label{tab:bm1} $M1$ transition strengths in 
$^{11}$B. 
The theoretical values were calculated
by the AMD (VAP) method. 
%The adopted interaction consists of
%the central force of MV1 ($m=0.62$, $b=h=0.25$),
%the spin-orbit force of GSRS ($u_I=-u_{II}=2800$ MeV), and Coulomb force. 
}{
\begin{tabular}{@{\qquad}c@{\qquad}c@{\qquad\qquad}c@{\qquad}c@{\qquad\qquad}}
%\toprule
%\begin{tabular}{@{}ccccc@{}} \toprule
%{ccccc}
%hline\hline
&   & \multicolumn{2}{c}{$B(M1;J_i\rightarrow J_f)$ $\mu_N^2$}   \\ 
$J_i$&$J_f$ & {exp.} & {theor.} \\
%\colrule
\hline
$1/2^-_1$&$3/2^-_1$	&1.07 (0.07) 	&	1.2 \\
$5/2^-_1$&$3/2^-_1$	&0.52 (0.02) 	&	0.72 \\
$3/2^-_2$&$3/2^-_1$	&1.13 (0.04) 	&	1.2 \\
$3/2^-_2$&$1/2^-_1$	&0.98 (0.04) 	&	1.0 \\
$7/2^-_1$&$5/2^-_1$	&0.006 (0.002) 	&	0.03 \\
$3/2^-_3$&$3/2^-_1$	&0.072 (0.007) 	&	0.07 \\
$3/2^-_3$&$1/2^-_1$	&0.091 (0.009) 	&	0.16 \\
$3/2^-_3$&$5/2^-_1$	&0.057 (0.013) 	&	0.04 \\
$3/2^-_3$&$3/2^-_2$	&0.163 (0.023) 	&	0.28 \\
$5/2^-_2$&$3/2^-_1$	&0.50 (0.02) 	&	0.45 \\
$5/2^-_2$&$5/2^-_1$	&0.21 (0.02) 	&	0.04 \\
\end{tabular}}
\end{table}

\begin{table}[ht]
\caption{ 
\protect\label{tab:be2} The quadrupole transition strengths 
in $^{11}$B.
The present results of the $B(E2)$, $M_p$ and $M_n$ are shown with the 
experimental values of $B(E2)$\protect\cite{Ajzenberg90}.
%The adopted interaction consists of
%the central force of MV1 ($m=0.62$, $b=h=0.25$),
%the spin-orbit force of GSRS ($u_I=-u_{II}=2800$ MeV), and Coulomb force. 
}{
\begin{tabular}{@{\qquad}c@{\qquad}c@{\qquad\qquad}c@{\qquad}c@{\qquad}c@{\qquad}c@{\qquad}}
%\toprule
%\begin{tabular}{@{}ccccc@{}} \toprule
%{ccccc}
%hline\hline
$J_i$&$J_f$ & {exp.} & {theor.} \\
&   & $B(E2)$ $e^2$ fm$^4$ & $B(E2)$ $e^2$ fm$^4$ 
& $M_p$ $e$ fm$^2$& $M_n$  $e$ fm$^2$\\ 
%\colrule
\hline
$1/2^-_1$	&	$3/2^-_1$	&		&	4.5 	&	3.0 	&	4.0 	\\
$5/2^-_1$	&	$3/2^-_1$	&	14(3)	&	12.8 	&	8.8 	&	7.5 	\\
$3/2^-_2$	&	$3/2^-_1$	&		&	0.0 	&	0.3 	&	2.7 	\\
$7/2^-_1$	&	$3/2^-_1$	&	1.9(0.4)	&	1.8 	&	3.8 	&	7.9 	\\
$3/2^-_3$	&	$3/2^-_1$	&		&	0.8 	&	1.8 	&	2.8 	\\
$5/2^-_2$	&	$3/2^-_1$	&	1.0(0.7)	&	0.1 	&	0.8 	&	0.9 	\\
$5/2^-_3$	&	$3/2^-_1$	&		&	0.7 	&	2.1 	&	1.2 	\\
\end{tabular}}
\end{table}

\section{Discussions} \label{sec:discuss}

In the present calculations of $^{11}$B and $^{11}$C, we found that
the $3/2^-_3$ states are the well developed three-center cluster states like
$2\alpha+t$ and $2\alpha+^3$He. We consider that these states are 
the candidates of the cluster gas state, which has an analogy 
to the $3\alpha$ gas state proposed in the $^{12}$C($0^+_2$).
On the other hand, the $5/2^-_2$ at almost the same excitation energy
as the $3/2^-_3$ is the non-cluster state.
In this section, we theoretically investigate the structure of 
$^{11}$B while focusing on cluster aspect, and show
the analogy of the excited states of $^{11}$B with those of 
$^{12}$C. 

\subsection{Intrinsic structure}
As explained in \ref{sec:formulation}, by performing the
VAP calculations
we obtained the optimum intrinsic states,
$\Phi_{\rm AMD}({\bf Z}^{J\pi}_n)$.
Although the final wave function $|J^\pi_n\rangle$ 
is expressed by the superposition of all the obtained AMD 
wave functions as \ref{eq:diago}, 
the spin-parity eigen state 
$|P^{J\pi}_{MK}\Phi_{\rm AMD}({\bf Z}^{J\pi}_{n})\rangle$ 
projected from the single AMD wave function is the dominant component of 
the $|J^\pi_n\rangle$ with the amplitude of more than $70\%$ 
in most cases except for the $3/2^-_3$. In the case $3/2^-_3$, 
since the amplitude is distributed into various AMD wave functions,
the amplitude of the dominant component
$|P^{3/2-}_{MK}\Phi_{\rm AMD}({\bf Z}^{3/2-}_{3})\rangle$
in the $|3/2^-_3\rangle$ is reduced to $50\%$.
Here we regard the obtained $\Phi_{\rm AMD}({\bf Z}^{J\pi}_n)$
written by the single Slater determinant as the approximate
intrinsic state of the corresponding $J^\pi_n$ state, and 
discuss the intrinsic structure.

\begin{figure}[th]
\epsfxsize=6 cm
\centerline{\epsffile{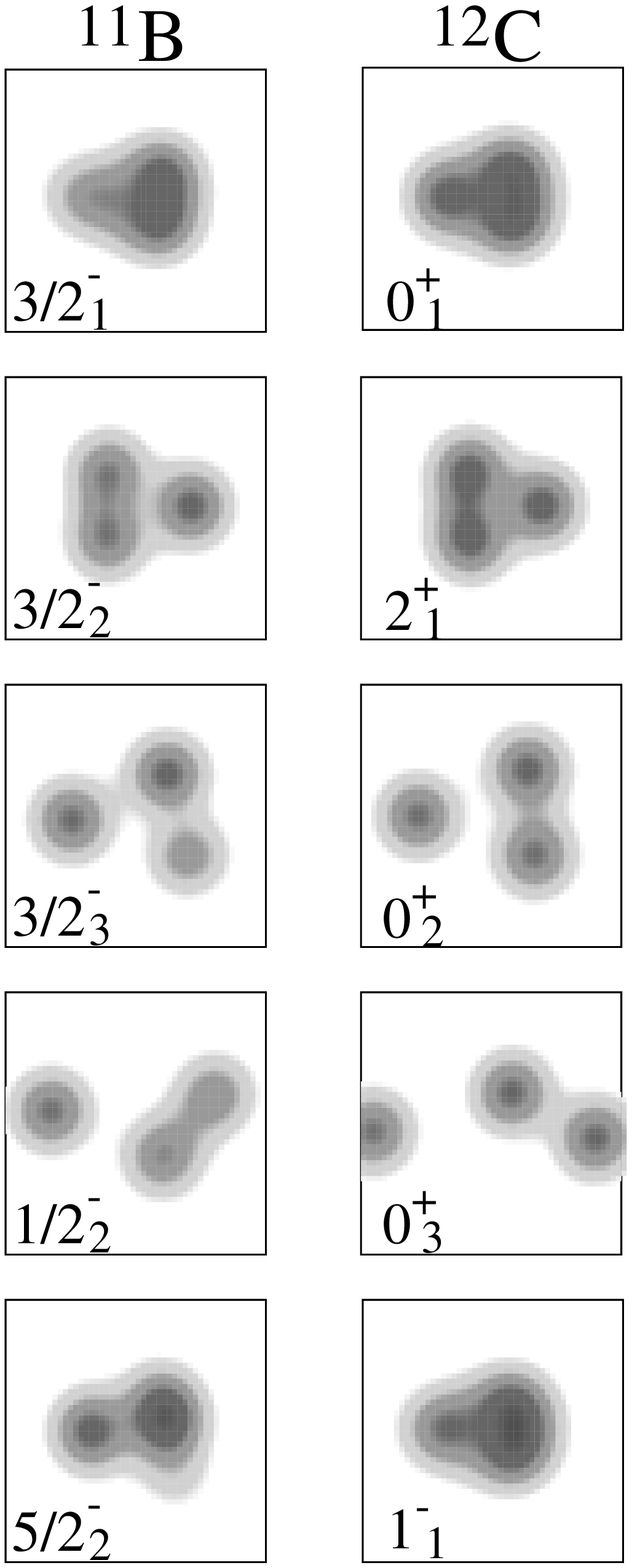}}
\vspace*{8pt}
\caption{Density distribution of the ground and excited states in
 $^{11}$B and $^{12}$C.
The density of the dominant AMD wave function of each state is shown.
\protect\label{fig:dense}}
\end{figure}

In Fig.~\ref{fig:dense}, we display the density distribution of 
the excited states of $^{11}$B. 
The matter density of the 
intrinsic wave functions $\Phi_{\rm AMD}({\bf Z}^{J\pi}_n)$
is shown.  
As shown in the density, the ground state($3/2^-_1$) has 
no cluster structure, while the
$3/2^-_2$ state has a structure with cluster cores. Since the 
spatial development of the clustering is not remarkable, 
the $3/2^-_2$ state is considered to be the $SU(3)$-limit cluster state.
Most striking thing is that the spatially developed cluster structure of 
$2\alpha+t$ appears in the $3/2^-_3$ state.
On the other hand, the $5/2^-_2$ state has no cluster structure 
though this state appears at almost the same excitation energy as the
$3/2^-_3$ state with the developed cluster structure.
In higher excited state, we found a somewhat 
linear-like $2\alpha+t$ cluster structure in the $1/2^-_2$.
The predicted $1/2^-_2$ state should be assigned to a 
$1/2^-,T=1/2$ state, however, the corresponding state 
has not been observed yet.

Let us show similarities of the cluster features seen 
in the intrinsic structure of  $^{11}$B with
those of $^{12}$C. We compare the present AMD results with 
those of $^{12}$C in Ref.\cite{Enyo-c12v2}. Then
we find a good correspondence of the intrinsic structure 
between $^{11}$B and $^{12}$C. As shown in Fig.\ref{fig:dense},
the ground states in both nuclei has no
remarkable cluster structure due to the nature of the $p_{3/2}$ 
sub-shell closure. The cluster core structure in the 
$^{11}$B$(3/2^-_2)$ state
is similar to that of the $^{12}$C$(2^+_1)$. 
Both the states show the three-center cluster core structure but the
spatial development is not remarkable. It means that the $^{11}$B$(3/2^-_2)$
and the $^{12}$C$(2^+_1)$ can be practically dominated by 
the $SU(3)$-limit cluster states of $2\alpha+t$ and $3\alpha$,
respectively.
The spatially developed $2\alpha+t$ clustering in the $^{11}$B$(3/2^-_3)$
is similar to the developed $3\alpha$ clustering in the $^{12}$C$(0^+_2)$. 
The details are discussed later.
The linear-like structure in the $^{11}$B$(1/2^-_2)$ is associated with
that of the $^{12}$C$(0^+_3)$ and $^{12}$C$(1^-_1)$ states. 
Although the structure of the $^{12}$C$(0^+_3)$ is not experimentally and
theoretically clarified yet, 
the linear-like $3\alpha$ structure in $^{12}$C was predicted by the 
generator coordinate method(GCM) calculation\cite{GCM} and
 also by the fermionic molecular 
dynamics\cite{Neff-c12} as well as the AMD.  
The $5/2^-_2$ has no cluster structure because this state appears due to
the intrinsic spin excitation, which causes the breaking of clusters. 
The situation is similar to the case of $^{12}$C$(1^+_1)$.

As mentioned before, the $^{11}$C$(3/2^-_3$, 8.10 MeV) 
and $^{11}$B$(3/2^-_3$, 8.65 MeV) have the abnormally 
small $B({\rm GT})$ and $B(M1)$ compared with the other low-lying states
in $E_x\le 9$ MeV.
The quenching of GT and $M1$ transitions 
for the $3/2^-_3$ states can be described by the above-mentioned 
exotic structure.
Namely, since the $^{11}$C($3/2^-_3$) and the $^{11}$B($3/2^-_3$) 
exhibit the well-developed $2\alpha+^3$He and $2\alpha+t$ 
clustering, they have small transition overlap with the 
other normal low-lying states.

\subsection{Dilute cluster states 
in the $3/2^-_3$}

By analyzing the obtained wave functions,
we found that the $^{11}$B$(3/2^-_3)$ is a 
three-center cluster state with the spatially developed $2\alpha+t$
clustering. 
The clustering features of the $^{11}$B$(3/2^-_3)$ and 
the $^{11}$C$(3/2^-_3)$ are 
very similar to those of 
$^{12}$C$(0^+_2$, 7.65 MeV), which is known to be 
a dilute gas-like $3\alpha$ state. Therefore, we consider that the 
$^{11}$C$(3/2^-_3$, 8.10 MeV) and the $^{11}$B$(3/2^-_3$,8.56 MeV) are 
the candidates of the dilute gas-like cluster states with 
$2\alpha+^3$He and $2\alpha+t$, respectively.
The similarity of the $^{11}$B($3/2^-_3$) with the $^{12}$C($0^+_2$) 
has been suggested in Ref.\cite{Kawabata05}, where the experimental 
data of the $(d,d')$ scattering have been analyzed.
We here theoretically discuss the 
similarity between the $^{11}$B($3/2^-_3$)  
and the $^{12}$C$(0^+_2)$ by comparing the 
wave functions of $^{11}$B and those of $^{12}$C obtained by 
the AMD as same as the present work\cite{Enyo-c12v2}.

\begin{figure}[th]
\epsfxsize=10. cm
\centerline{\epsffile{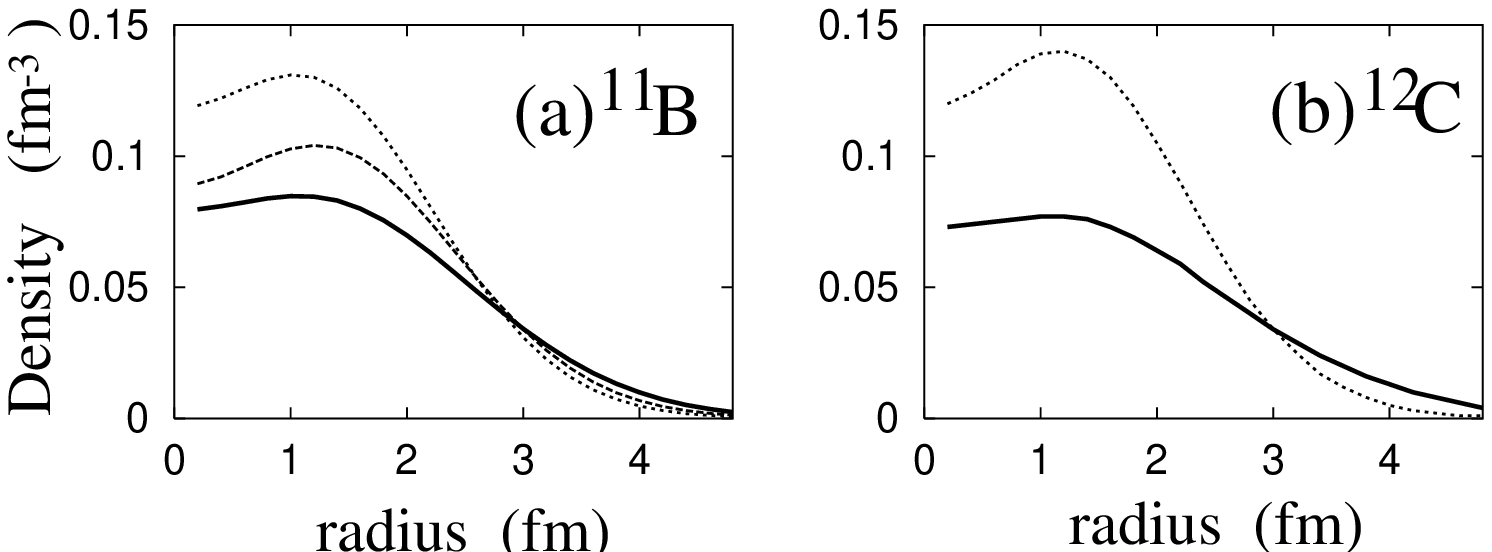}}
\vspace*{8pt}
\caption{Point matter density of 
(a) the $3/2^-_1$(dotted), $3/2^-_2$(dashed) and $3/2^-_3$(solid)
 states of $^{11}$B, and 
(b) the $0^+_1$(dotted) and $0^+_2$(solid) of $^{12}$C.
\protect\label{fig:rdens}}
\end{figure}

In order to see the diluteness of the cluster states, 
first we plot the matter density $\rho(r)$ as a function of the radius $r$
in Fig.~\ref{fig:rdens}.
In the ground states of $^{11}$B 
and $^{12}$C, the density distributes in the small $r$ region because of their
compact structures. On the other hand, the density in the 
$^{11}$B($3/2^-_3$) state is about a half of the normal density 
at the center and
has a tail in the outer region due to the spatial development of clusters.
The density curve of the $^{11}$B($3/2^-_3$) is similar to that of the
$^{12}$C($0^+_2$) though the outer tail is less remarkable than the
$^{12}$C($0^+_2$).
Next we show 
the matter root-mean-square radii of the ground and the excited states of
$^{11}$B and $^{12}$C in Table \ref{tab:rmsr}. 
In $^{12}$C, the $0^+_2$ has a large radius. The calculated value of the 
$0^+_2$ is 3.3 fm in the AMD calculations, while those 
obtained by the RGM calculations\cite{RGM} 
and the $\alpha$ condensate wave functions\cite{Funaki03} are 
3.5 fm and 3.8 fm, respectively.
The smaller theoretical radius in the present method is considered to 
be because of the limited number of the base wave functions.
In $^{11}$B, the radius of the $3/2^-_3$ state is remarkably large compared
with the size of the ground state. Considering the large radius and
the density tail in the outer region,
we can say that the $3/2^-_3$ state shows the nature of a dilute cluster state.

\begin{table}[ht]
\caption{ 
\protect\label{tab:rmsr} 
Matter root-mean-square radii(r.m.s.r.) and
expectation values of the harmonic oscillator quanta for 
protons($\Delta Q_p$) and neutrons($\Delta Q_n$).
The values of $\Delta Q$ are defined by subtracting the minimum 
oscillator quanta. See the details in the text.  The expectation
values of the squared intrinsic spin for neutrons 
$\langle {\bf S}^2_n \rangle$ are also listed.
The observed r.m.s.r. of the $^{12}$C($0^+_1$) is estimated to be 
$2.32-2.33$ fm by the electron scattering data.
}{
\begin{tabular}{@{\qquad}c@{\qquad\qquad}c@{\qquad}c@{\qquad}c@{\qquad}c}
% \toprule
%{ccccc}
%hline\hline
           & r.m.s.r. (fm) & $\Delta Q_p$ & $\Delta Q_n$ 
& $\langle {\bf S}^2_n \rangle$\\
%\colrule
$^{11}$B($3/2^-_1$) &  2.5 & 0.3 & 0.4 & 0.7\\
$^{11}$B($3/2^-_2$) &  2.7 & 0.9 & 1.1 & 0.2\\
$^{11}$B($3/2^-_3$) &  3.0 & 2.0 & 2.6 & 0.4 \\
$^{11}$B($1/2^-_1$) &  2.7 & 0.7 & 0.8 & 0.3 \\
$^{11}$B($1/2^-_2$) &  3.1 & 3.0 & 3.6 & 0.3 \\
$^{11}$B($5/2^-_1$) &  2.6 & 0.5 & 0.7 & 0.5\\
$^{11}$B($5/2^-_2$) &  2.6 & 0.5 & 0.7 & 1.3 \\
$^{11}$B($5/2^-_3$) &  2.7 & 0.7 & 0.9 & 0.8 \\
\hline
$^{12}$C($0^+_1$) &  2.5 & 0.4 & 0.4 & 0.6\\
$^{12}$C($0^+_2$) &  3.3 & 4.4 & 4.3 & 0.3 \\
$^{12}$C($0^+_3$) &  4.0 & 10.0 & 9.9 & 0.1 \\
$^{12}$C($2^+_1$) &  2.7 & 0.8 & 0.8 & 0.2 \\
$^{12}$C($1^+_1$) &  2.5 & 0.2 & 0.2 & 1.4 
\end{tabular}}
\end{table}

In order to give more quantitative discussions of 
the spatial development of clusters, 
we examine the expectation values of the harmonic oscillator(H.O.) quanta for 
protons and neutrons in table \ref{tab:rmsr}. 
For the width parameters of the H.O., 
we use the same width of the Gaussian wave packets adopted in the
AMD wave function. The values $\Delta Q$ 
are defined by subtracting the minimum 
oscillator quanta from the expectation values of the principal quantum number
of H.O.,
\begin{equation}
\Delta Q\equiv \langle a^\dagger a\rangle-Q_{\rm min},
\end{equation}
where $Q_{\rm min}$ is 3(4) and 4(4) for protons(neutrons) of 
$^{11}$B and $^{12}$C, respectively.
The expectation values of the oscillator quanta indicate the higher shell
components in terms of the H.O. shell model. 
It is generally enhanced when the clustering spatially 
develops, because it  
necessarily increases the higher shell components. 
In the $^{12}$C($0^+_2$) and the $^{11}$B($3/2^-_3$), the 
large $\Delta Q$ values are caused by the developed three-center
cluster structure. 
Such the higher shell components due to the cluster correlation
in the developed cluster states can not be treated 
in the truncated space of shell model. This is the reason why
the shell model calculations fail to describe the 
$^{12}$C($0^+_2$) and the $^{11}$B($3/2^-_3$).
On the other hand, the $\Delta Q$ values 
in the $^{11}$B($3/2^-_2$) are rather small compared with 
those of the $3/2^-_3$.
It means that the major component of the 
$3/2^-_2$ is the $0\hbar\omega$ configuration.
Since it has a compact state with cluster cores 
as shown in Fig.~\ref{fig:dense}, 
this state is interpreted to be almost the SU(3)-limit cluster state.

\begin{figure}[th]
\epsfxsize=10. cm
\centerline{\epsffile{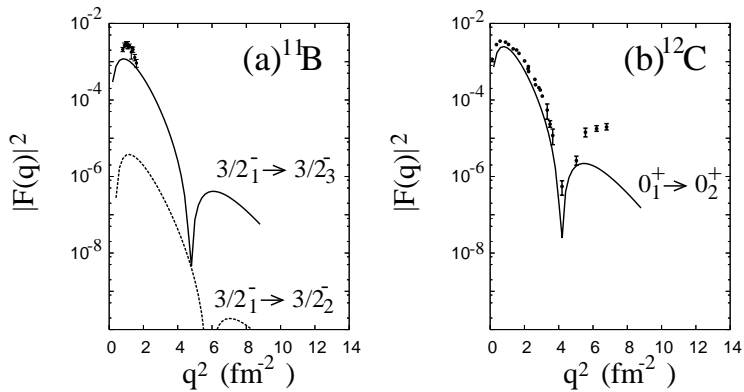}}
\vspace*{8pt}
\caption{
Squared inelastic form factors 
for the electron scattering on (a)$^{11}$B and (b)$^{12}$C.
The experimental form factors for the transitions to 
$^{11}{\rm B}(3/2^-,8.56 {\rm MeV})$ and 
$^{12}$C$(0^+_2)$ are taken from \protect\cite{Polishchuk79} 
and \protect\cite{Sick70}, respectively.
The lines are the calculated form factors of the $E0$ components. 
\protect\label{fig:form}}
\end{figure}

The similarity between 
the $^{11}$B($3/2^-_3$) and the $^{12}$C($0^+_2$) 
has been suggested in Ref.\cite{Kawabata04,Kawabata05}, where the multipole
decomposition analysis of the inelastic 
$(d,d')$ scattering has been performed. The remarkable strengths of inelastic 
monopole transitions are
the characteristics of these states. Figure \ref{fig:form}.
shows the calculated electron
form factor for the monopole transitions, 
$^{12}$C($0^+_1\rightarrow 0^+_2$),
$^{11}$B($3/2^-_1\rightarrow 3/2^-_2$) and  
$^{11}$B($3/2^-_1\rightarrow 3/2^-_3$).
The profile and the absolute value of the form factor are
similar between the $^{11}$B($3/2^-_1\rightarrow 3/2^-_3$) and
the $^{12}$C($0^+_1\rightarrow 0^+_2$), while the 
form factor for the $^{11}$B($3/2^-_1\rightarrow 3/2^-_2$) is 
more than factor $10^2$ smaller. This is consistent with the
experimental results of the $(d,d')$ scattering\cite{Kawabata04,Kawabata05}.

As mentioned above, we can see the developed cluster structure 
with dilute density in the $^{11}$B$(3/2^-_3)$ as well as the
$^{12}$C$(0^+_2)$. 
The $^{12}$C$(0^+_2)$ is interpreted as 
a cluster gas state, where 3 $\alpha$ clusters are rather freely moving\cite{Tohsaki01,Funaki03}.
Here ``a cluster gas'' means the well developed cluster state with 
dilute density, where the clusters are freely moving in terms of the 
weak coupling picture.
Such the gas-like nature is reflected not only in the dilute density but
also in the fragmentation of the amplitudes in the AMD model space.
Let us remind the reader that a base AMD wave function 
is expressed by a Slater determinant.
If the cluster state is written by an AMD wave function, it has a certain
spatial configuration of the cluster centers like a 
single Brink-type cluster wave function\cite{Brink66}.
On the contrary, 
when the state has a cluster gas-like
feature, its wave function is written by a superposition of various 
AMD wave functions with different configurations of cluster centers. 
As a results, the cluster gas state is not dominated 
by a single AMD wave function, but
the amplitudes distribute in various base wave functions.
Actually in the $^{12}$C($0^+_2$), the amplitude of the 
$|P^{0+}_{MK}\Phi_{\rm AMD}({\bf Z}^{0+}_{2})\rangle$ 
is reduced to about $50\%$ because of the cluster gas nature 
as discussed in Ref.\cite{Enyo-c12v2}.
Similarly, in the case of $^{11}$B$(3/2^-_3)$, the amplitude of the
dominant component is only $50\%$, while those for the 
$^{11}$B$(3/2^-_1)$ and $^{11}$B$(3/2^-_2)$ are
more than $90\%$. This indicates the gas-like nature of $2\alpha+t$ cluster
in the $^{11}$B$(3/2^-_3)$ as well as the $3\alpha$ cluster 
in the $^{12}$C$(0^+_2)$.

Considering the smaller radius of the 
$^{11}$B($3/2^-_3$) than the $^{12}$C$(0^+_2)$, the cluster gas-like nature
in the $^{11}$B($3/2^-_3$) is not so remarkable as that in 
the $^{12}$C$(0^+_2)$. We consider the reasons for the 
less gas-like nature in the $^{11}$B($3/2^-_3$) as follows.
Firstly, the inter-cluster potential is more attractive 
in the $\alpha-t$ channel than the $\alpha-\alpha$ channel.
This is already known in the comparison of the binding energy 
between $^7$Li and $^8$Be. The origin is that 
the repulsive Pauli effect is smaller in the $\alpha-t$ 
than the $\alpha-\alpha$.
Second, from the natural extension of the 
the ground state properties of $^7$Li and $^8$Be 
it is expected that the triton motion may have the 
orbital angular-momentum $L=1$ while the motion of the $\alpha$ 
clusters has $L=0$. The $L=1$ should be less favored to form a dilute cluster gas 
state than the $L=0$. Thirdly, it might be important that 
the symmetry of three clusters 
is not good in the $2\alpha+t$ system as the $3\alpha$ system. Because of the
symmetry of 3 $\alpha$ orbits, the $^{12}$C($0^+_2$)
state is understood as the $\alpha$ condensate state 
as argued in Refs.\cite{Tohsaki01,Funaki03,Yamada04,Matsumura04}. However, 
it is not easy to define the Bosonic behavior and to
discuss the condensation in the $2\alpha+t$ system, which
contains only two identical bosons.

In the stabilizing mechanism of the dilute cluster states, 
one of the keys preventing the states from
shrinking is the orthogonality to the compact states 
in the lower energy region. In both cases of $^{12}$C and $^{11}$B, 
there exist the lower states with the compact cluster components. 
In the higher cluster states, the cluster distribution avoids 
the compact inner region and must spread out to satisfy the 
orthogonality to the lower states. It is interesting that the
the number of the lower compact states is one($0^+_1$) in $^{12}$C 
and it is two($3/2^-_1$,$3/2^-_2$) 
in case of $^{11}$B, which is just the number of
the low-lying states described 
by the $0\hbar\omega$ configurations.
This is the reason why the dilute cluster state
appears in the {\it third} $3/2^-$ state
in the $^{11}$B system.

The diluteness of the cluster states should be sensitive also 
to the relative energy against the threshold energy of the corresponding
cluster channel.
In the present results, the threshold energy for the three-body cluster
breakup is not reproduced. In order to check the dependence 
of the relative energy of the excited state to the threshold, 
we vary the relative energy by changing the interaction parameters, 
and found that the structure of the excited states  
are qualitatively unchanged. It means that the present results 
are not sensitive to 
the relative position of the threshold.  
It is because the present framework is a kind of bound state
approximation.
Since the number of the base wave functions is limited, 
the long tail part of the inter-cluster motion may not be enough 
taken into account, and therefore  
the description of the detailed resonant behavior is
insufficient in the present framework. 
In fact, the AMD calculations give the smaller 
radius of the $^{12}$C$(0^+_2)$ than that obtained by 
the $3\alpha$ cluster models.
However, we should stress that the features of $3\alpha$ system 
obtained by the present method are qualitatively 
similar to those of the $3\alpha$-GCM calculations
by Uegaki {\it et al.}\cite{GCM}, and also consistent 
with the $3\alpha$ model of orthogonal condition method(OCM) 
with the complex scaling method(CSM)\cite{Kurokawa04} where
the resonant behavior was appropriately treated. 
It implies that the present results are useful 
for qualitative discussions, though  
the long tail part of the inter-cluster motion and its 
boundary conditions should be carefully treated
for further quantitative study.

We should comment that the loosely bound cluster states have been 
predicted by Nishioka {\it et al.} with a $2\alpha+t$-OCM 
cluster model\cite{Nishioka79}.
However, they could not assign the $3/2^-_3$ state because the reproduction
of the energy spectra in the low-energy region was poor 
in the cluster model space.

\subsection{Intrinsic spin excitation}

In the ideal $2\alpha+t$ and $3\alpha$ cluster states,
the expectation values of the 
squared total intrinsic spin for neutrons($\langle {\bf S}^2_n\rangle$) 
should be zero, because spin-up and spin-down neutrons couple 
to be spin-zero pairs.
In $^{11}$B and $^{12}$C, non-zero values of $\langle {\bf S}^2_n\rangle$
is caused by the component of the cluster breaking.
We show the values of $\langle {\bf S}^2_n\rangle$ in 
$^{11}$B and also those in $^{12}$C in table \ref{tab:rmsr}. 
The $\langle {\bf S}^2_n\rangle$ values are small in the 
cluster states
such as the $^{11}$B$(3/2^-_2)$, $^{11}$B$(3/2^-_3)$ and 
$^{11}$B$(1/2^-_2)$.
while that of the $^{11}$B$_{\rm g.s.}$ is significantly large as
$\langle {\bf S}^2_n\rangle=0.7$ due to the component of the
$p_{3/2}$ sub-shell closure as well as that of the $^{12}$C$_{\rm g.s.}$.
An interesting point is the large value, $\langle {\bf S}^2_n\rangle=1.3$, in
the $^{11}$B($5/2^-_2$). This means that the $5/2^-_2$ state is 
characterized by the intrinsic spin excitation of neutrons within 
the $p$-shell.
This feature well corresponds to the structure of the $^{12}$C$(1^+_1)$,
which is assigned to the observed $1^+_1(12.7$ MeV) state.
In the comparison of the experimental excitation energies
between $^{11}$B($5/2^-_2$)
and $^{12}$C$(1^+_1)$ with the intrinsic spin excitation,  
it is interesting that the $^{11}$B($5/2^-_2, 8.92$ MeV) 
appears in the low-energy region and almost degenerate  
with the cluster gas-like $^{11}$B($3/2^-_3$, $8.56$ MeV), 
while the $^{12}$C$(1^+_1, 12.7$ MeV) exists at much higher 
excitation energy than the $^{12}$C$(0^+_2, 7.6$ MeV).
It implies that the intrinsic spin excitation easily occurs in
$^{11}$B than the $^{12}$C, and that 
the excitation energy of the intrinsic 
spin excitation is almost the same as that of the cluster 
excitation in $^{11}$B.
 
%\begin{figure}[th]
%\epsfxsize=10 cm
%\centerline{\epsffile{dense-mono.eps}}
%\vspace*{8pt}
%\caption{Density distribution of 
%the $3/2^-_1$, $3/2^-_2$ and $3/2^-_3$ states of $^{11}$B.
%The density of the dominant AMD wave function of each state is shown.
%\protect\label{fig1}}
%\end{figure}

\section{Summary}\label{sec:summary}
We studied the negative-parity states 
in $^{11}$B and $^{11}$C based on the theoretical calculations of
antisymmetrized molecular dynamics(AMD). It is concluded that
various types of cluster states appear in the
excited states of $^{11}$B and $^{11}$C. 
Recent experimental data of GT transition strengths for 
the $3/2^-_3$ and the $5/2^-_2$ states at $E_x\sim 8$ MeV
are well reproduced by the cluster state and the non-cluster state,
respectively.
It was found that the excitation energy of the intrinsic 
spin excitation is almost the same as that of the cluster 
excitation in $^{11}$B.
We compared the cluster aspect in the excited states of $^{11}$B with that of 
$^{12}$C, and showed a good similarity between the $2\alpha+t$ and $3\alpha$
systems.

We succeeded to describe the $^{11}$B($3/2^-_3,8.56$ MeV) and 
$^{11}$C($3/2^-_3,8.10$ MeV) states, 
which have not been reproduced by any other models.
For the assignment of the theoretical states to the observed ones, 
it is essential to systematically describe 
the properties of the coexisting cluster and non-cluster states 
in $^{11}$C and $^{11}$B.
One of the new revelations in the present
work is that $^{11}$C$(3/2^-_3$) and $^{11}$B$(3/2^-_3$)
are the well developed cluster states of 
$2\alpha+^3$He and $2\alpha+t$ with dilute density, respectively. 
The features of these dilute
cluster states in $^{11}$C and $^{11}$B are similar to those of the $0^+_2$
state of $^{12}$C, which is understood to be a cluster gas of weakly 
interacting 3$\alpha$ particles. 

Since the present framework is a kind of bound state approximation,
the description of resonant behavior is not
sufficient.
The boundary conditions of the inter-cluster motion
should be taken into account more carefully in more detailed 
investigations of the developed cluster states.

\section*{Acknowledgments}

The author would like to thank Prof. Schuck, Prof. Kawabata 
and the member of the
``Research Project for Study of
Unstable Nuclei from Nuclear Cluster Aspects'' sponsored by
Institute of Physical and Chemical Research (RIKEN) for many 
discussions.
The computational calculations in this work were performed by the 
Supercomputer Projects 
of High Energy Accelerator Research Organization(KEK)
and also the super computers of YITP.
This work was supported by Japan Society for the Promotion of 
Science and a Grant-in-Aid for Scientific Research of the Japan
Ministry of Education, Science and Culture.
The work was partially performed in the ``Research Project for Study of
Unstable Nuclei from Nuclear Cluster Aspects'' sponsored by
Institute of Physical and Chemical Research (RIKEN).

\end{document}